\documentclass{aa}  

\usepackage{graphicx}
\usepackage{txfonts}
\usepackage{subcaption}         % necessary for continued figures, example in section 3
\usepackage{lscape}             % to rotate a single page table, example in appendix.
\usepackage{placeins}           % useful with \FloatBarrier, to keep 
\usepackage[colorlinks=true, allcolors=blue]{hyperref}
\defcitealias{LimongiChieffi2018}{LC18}

\newcommand{\amnh}{Department of Astrophysics, American Museum of Natural History, 200 Central Park West, New York, NY 10024, USA}
\newcommand{\bath}{Department of Physics, University of Bath, Claverton Down, Bath, BA2 7AY, UK}
\newcommand{\surrey}{University of Surrey, Physics Department, Guildford, GU2 7XH, UK}
\newcommand{\columbia}{Department of Astronomy, Columbia University, New York, NY 10027, USA}
\newcommand{\uchicago}{Department of Astronomy \& Astrophysics, University of Chicago, 5640 S Ellis Ave, Chicago, IL 60637, USA}
\newcommand{\kicp}{Kavli Institute for Cosmological Physics, University of Chicago, Chicago, IL 60637, USA}
\newcommand{\lund}{Lund Observatory, Division of Astrophysics, Department of Physics, Lund University, Box 43, SE-221 00 Lund, Sweden}
\newcommand{\hertfordshire}{Centre for Astrophysics Research, University of Hertfordshire, Hatfield, AL10 9AB, UK}
\newcommand{\nsfsimons}{NSF-Simons AI Institute for the Sky (SkAI), 172 E. Chestnut St., Chicago, IL 60611, USA}
\newcommand{\cfa}{Center for Astrophysics | Harvard \& Smithsonian, Cambridge, MA 02138, USA}

\begin{document}

%%%%%%%%%%%%%%%%%%%%%%%%%%%%%%%%%%%%%%%%
% if you use custom commands in your title,
% ensure to check your title when submitting!
%%%%%%%%%%%%%%%%%%%%%%%%%%%%%%%%%%%%%%%%
   \title{EDGE-INFERNO: How chemical enrichment assumptions impact the individual stars of a simulated ultra-faint dwarf galaxy}
   \titlerunning{EDGE-INFERNO: Chemical enrichment in ultra-faint dwarf galaxies}
   
   \author{Eric~P.~Andersson\inst{1}
          \and Martin~P.~Rey\inst{2}
          \and Robert~M.~Yates\inst{3}
          \and Justin~I.~Read\inst{4}
          \and Oscar~Agertz\inst{5}
          \and Alexander~P.~Ji\inst{6,7,8}
          \and Jennifer~Mead\inst{9}
          \and Kaley~Brauer\inst{10}
          \and Mordecai-Mark~Mac~Low\inst{1}
          }
          
    \institute{\amnh\\
              \email{eandersson@amnh.org}
              \and \bath
              \and \hertfordshire
              \and \surrey
              \and \lund
              \and \uchicago
              \and \kicp
              \and \nsfsimons
              \and \columbia
              \and \cfa
             }

   \date{Received November 7, 2025}

  \abstract
  {The chemical abundances of stars in galaxies are a fossil record of the star formation and stellar evolution processes that regulate galaxy formation, including the stellar initial mass function, the fraction and timing of type Ia supernovae (SNeIa), and nucleosynthesis inside massive stars. In this paper, we systematically explore uncertainties associated with modeling chemical enrichment in dwarf galaxies. We repeatedly simulate a single \textsc{edge-inferno} dwarf ($M_{\star} \approx 10^5 \, M_{\odot}$), varying the chemical yields of massive stars, the timing and yields of SNeIa, and the intrinsic stochasticity that arises from sampling individual stars and galaxy formation chaoticity. All simulations are high-resolution ($3.6\,{\rm pc}$), cosmological zoom-in hydrodynamical simulations that track the stellar evolution of all individual stars with masses $>0.5\,{\rm M}_{\odot}$. We find that variations in SNIa assumptions make the largest difference in mean abundance ratios and [Fe/H], highlighting the importance of detailed SNIa modeling even in such low-mass reionization-limited galaxies. In contrast, different massive star yields, accounting (or not) for stellar rotation, result in mean abundances comparable to those arising from stochasticity. Nonetheless, they significantly affect the shape of abundance trends with [Fe/H], for example, through the existence (or not) of a bimodality in the [X/Fe] - [Fe/H] planes, particularly in [Al/Fe]. Finally, we find that the variance arising from random sampling severely limits the interpretation of single galaxies. Our analysis showcases the power of star-by-star cosmological models to unpick how both systematic uncertainties (e.g., assumptions in low-metallicity chemical enrichment) and statistical uncertainties (e.g., averaging over enough galaxies and stars within a galaxy) affect the interpretation of chemical observables in ultra-faint dwarf galaxies.
 }

   \maketitle
   
%%%%%%%%%%%%%%%%%%%%%%%%%%%%%%%%%%%%%%%%%%%%%%%%%%%%%%%%%%%%%%
\section{Introduction}\label{sec:introduction}

The faintest dwarf galaxies live in such shallow gravitational potential wells that they are expected to see their star formation histories abruptly suppressed at high redshift by reionization \citep{Efstathiou1992, Shapiro+1994, Gnedin2000, Bullock+2000, Benson+2002a, Benson+2002b, Somerville2002}. The population of ultra-faint dwarfs (UFDs; $M_V \gtrsim -7.75$; $L \leq 10^5 L_{\odot}$; see \citealt{Simon2019} for a review), with most of their stars formed at $z\geq 4$ (\citealt{Brown+2014, Weisz+2014, Savino+2023, Savino+2025a, Durbin+2025}), is a natural place to study such low-mass, reionization-limited galaxies.

In particular, the chemical elements locked in the photosphere of low-mass, long-lived stars in these dwarfs encode a fossil record of star formation events at $z\geq 4$. Such chemical abundances can often be linked to specific sites of nucleosynthesis (\citealt{Tinsley+1980, Timmes+1995, Matteucci2001}, see also \citealt{Maiolino&Mannucci2019, Kobayashi+2020} for recent reviews), in turn providing us with a unique window into star formation, stellar evolution, and nucleosynthesis processes that are hard to probe directly. Furthermore, measurements of chemical abundances in stars within faint dwarfs has been steadily increasing \citep[see][for recently compiled data]{Pace2024}, in both [Fe/H] measurements \citep[e.g.,][]{Munoz2006, Martin+2007, Simon&Geha2007, Fu+2023, Luna+2025, Sandford+2025} and other abundance ratios \citep[e.g.,][]{Kirby+2009, Frebel+2010, Norris+2010, Vargas+2013, Ji+2020, Skuladottir+2017, Skuladottir+2024}.

These measurements have strongly established that the positive correlation between galaxy luminosity and stellar [Fe/H] (closely related to the stellar mass-metallicity relation) extends well into the regime of faint galaxies \citep{Simon&Geha2007, Kirby+2013, Read+2017}, albeit with a possible flattening and/or increased scatter at V-band magnitude $M_{\rm V}\gtrsim-5.5$ \citep[see, e.g.,][]{Simon2019, Fu+2023}. Both the slope, the normalization, the scatter, and the potential plateau in this relation are sensitive to the details of how stellar feedback drives outflows in shallow potential wells \citep{Agertz+2020, Rey+2025}, to the stellar initial mass function \citep[IMF,][]{Prgomet+2022}, to the formation of the first metal-free Population (Pop.) III stars \citep{Jeon+2017, Gutcke+2022, Sanati+2023, Rey+2025b}, and to the specific galaxy growth history \citep{Rey+2019}. As such, this relation is
often used to compare and discriminate between different dwarf galaxy formation models (e.g. \citealt{Ko+2025, Rey+2025b, Rey+2025, Wheeler+2025}).

Beyond the average iron content and its relation with stellar mass, one can also leverage the distribution of individual metals (hereafter metal distribution functions, MDFs), their relationship to one another in fractional abundances, and to [Fe/H] as a tracer of total metallicity. For example, a downturn in the [$\alpha$/Fe]-[Fe/H] plane (where $\alpha$ represents one or the sum of e.g. O, Mg, Si, Ca, Ti) is usually interpreted as revealing the time at which SNeIa start contributing significantly to iron production compared to massive stars. Such a "knee" is detected in many, but not all, faint dwarfs \citep{Frebel+2012, Frebel+2014, Hill+2019, Chiti+2023} which could be leveraged into constraints on star formation and feedback timescales (e.g. \citealt{Kirby+2019, Alexander+2025, Ting&Ji2025}. Similarly, the shape of the [Fe/H] MDF can be modeled through semi-analytical models to determine the length of the star formation histories and the strength of outflows (e.g. \citealt{Alexander+2023, Sandford+2024, Luna+2025}), while specific abundances in UFDs have been linked to properties of metal-free stars (e.g. \citealt{Jeon+2017, Rossi+2021}). 

However, the interpretation of chemical abundance data in dwarf galaxies relies on our understanding of how enrichment from stars formed at $z\geq 4$ is imprinted into present-day observables. Predictions of elemental abundances in simulations necessitate chemical yields associated with different nucleosynthesis sites. Deriving these yields is computationally expensive, and instead, models of galactic chemical evolution rely on pre-tabulated estimates \citep[see, e.g.,][]{Cote+2013, Cote+2016, Kobayashi+2020, Buck+2021, Buck+2025}, even for semi-analytical modeling \citep{Kobayashi+2020, Yates+2021, Spitoni+2023}. As such, pre-tabulated yields have been available for a long time and are frequently being updated. These are typically divided into different sources, e.g., asymptotic giant branch (AGB) stars \citep[e.g.,][]{Marigo+2001, Karakas2010, Karakas&Lugaro2016, Pignatari+2016, Ritter+2018}, super-AGB and electron capture SNe \citep{Siess2010, Doherty+2014a, Doherty+2014b, Gil-Pons+2022, Limongi+2024, Limongi+2025}, core-collapse SNe (CCSNe) and winds from massive stars \citep[e.g.,][]{Raiteri+1996, Portinari+1998, Francois+2004, Chieffi&Limongi2004, Nomoto+2013, Frischknecht+2016, Pignatari+2016, LimongiChieffi2018, Ritter+2018}, SNeIa \citep[e.g.,][]{Iwamoto+1999, Thielemann+2003, Seitenzahl+2013}, hypernovae \citep{Umeda&Nomoto2002, Umeda+2002, Nomoto+2013}, and pair-instability SNe \citep{Heger&Woosley2002, Umeda&Nomoto2002, Cooke&Madau2014}. 

However, even when focusing on the same stellar types, yields carry significant systematic uncertainties \citep[e.g., stellar rotation, magnetic field strength, nuclear reaction rates, and modeling of explosions,][]{Romano+2010, Seitenzahl+2013, Cote+2016, Philcox+2018, Kobayashi+2020b}, often resulting in significant differences in the chemical content of galaxies \citep[][]{Buck+2021}. These uncertainties are amplified at the low metallicities of UFDs, for which few massive stars can be observed around the Milky Way to anchor stellar evolution tracks. Because of these uncertainties, there is a pressing need to explore their impact on predictions of the chemical content of dwarf galaxies and, as a result, the ability of the chemistry of UFDs to shed light on metal-free and metal-poor stellar evolutionary processes. 

In this work, we conduct such a systematic and controlled exploration to quantify how the abundance ratio of an ultra-faint dwarf responds to changes in chemical yields and chemical enrichment assumptions. We use high-resolution cosmological simulations to self-consistently follow the dark matter assembly, star formation, and chemical enrichment history of a single UFD. 

All simulations use a detailed stellar feedback and enrichment model, \textsc{inferno}, which incorporates chemical ejecta star by star rather than relying on integrated stellar populations \citep{Andersson+2023, Andersson+2025}. The model ties the release of selected elements to individual stars going through different stages of evolution, including winds from massive O and B-type stars on the main sequence, winds from stars on the giant branch, and SNe, both as a result of CCSNe and from binary evolution (SNeIa). For each source, nucleosynthesis yields in the ejecta were calculated using bilinear interpolation of pre-tabulated values obtained in the literature. The model is flexible, and any yields provided for the sources tracked in our model can be incorporated with only minor adjustments to the code base. With this tool at hand, we explore sets of commonly applied tables found in the literature \citep{Pignatari+2016, Ritter+2018, LimongiChieffi2018, Seitenzahl+2013}, aiming to test the robustness of abundance signatures, MDFs, and average [Fe/H] across assumptions.

We describe the numerical methods in Section~\ref{sec:method}, including our star-by-star enrichment model in \ref{sec:yield_method} and the cosmological simulation setup in Section~\ref{sec:numerical_method}. We present the average iron content, MDFs, and abundances in 13 re-simulations of the same object in Section~\ref{sec:result}, varying the yields of massive stars, the parametrization and yields of SNeIa, and the compounded stochastic variance from sampling star formation and feedback using random number generators. We discuss our results and the limitations of our model in Section~\ref{sec:discussion}, and conclude in Section~\ref{sec:conclusion}.

\section{Methods}\label{sec:method}

Our model works on a star-by-star basis. In this section, we describe our treatment of stellar evolution and how stars interact with their surroundings through feedback in Section~\ref{sec:yield_method}. We describe the numerical treatment of star formation, IMF sampling, stellar dynamics, and feedback injection in Section~\ref{sec:numerical_method}. The total mass loss for any source is based on look-up tables, described in Section~\ref{subsec:yield_tables}.

%----------------------------
\subsection{Star-by-star chemical enrichment}\label{sec:yield_method}
%----------------------------

\begin{figure*}
    \centering
    \includegraphics[width=\linewidth]{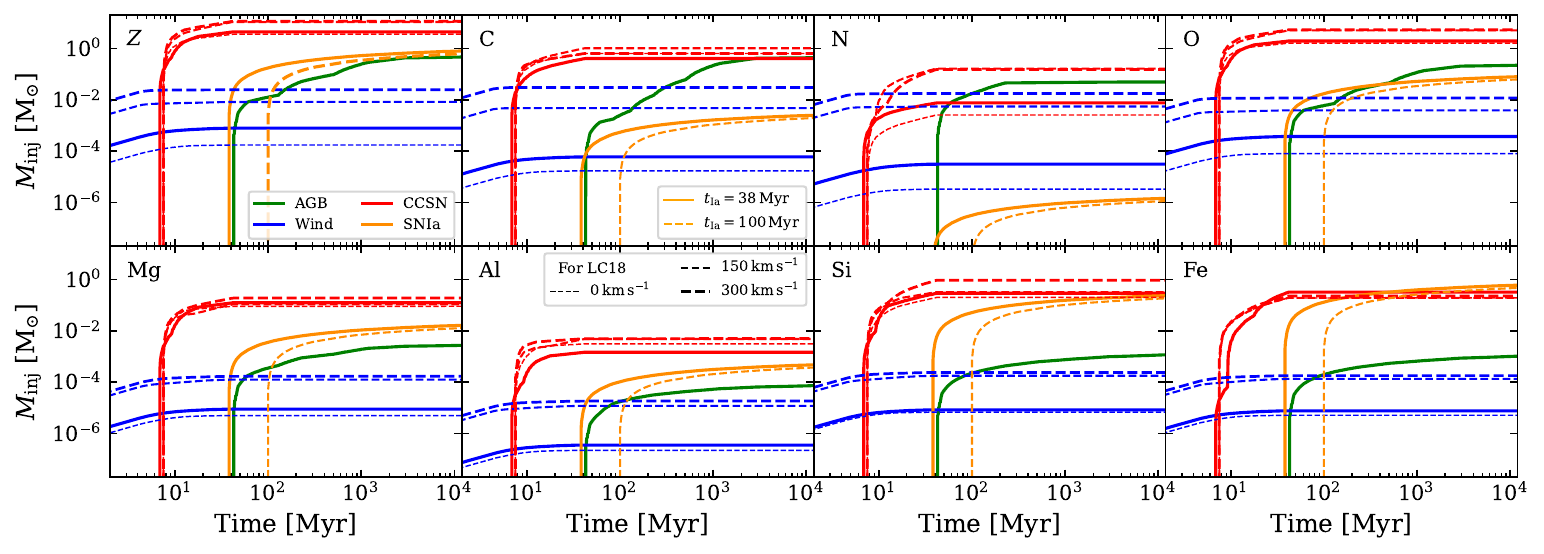}
    \caption{The average cumulative mass of different elements (and total metallicity $Z$) injected by a mono-age population with mass $500\,{\rm M}_{\odot}$. Colors indicate different sources as labeled in the legend of the upper left panel. Except for SNIa, filled lines show the yields from \textsc{NuGrid}, while dashed lines show those from \citet{LimongiChieffi2018} with different models ($0,150,300\,{\rm km\,s}^{-1}$) shown with thicker lines for increasing rotation velocity. For SNIa, the filled (dashed) line shows the result with a delay time of $38\,{\rm Myr}$ ($100\,{\rm Myr}$).}
    \label{fig:yields}
\end{figure*}

The model approximates stellar evolution with a few important stellar phases. We omit proto-stellar evolution, starting stars on the main sequence for a duration calculated based on the mass and metallicity dependent fitting function by \citet{Raiteri+1996}. After the main-sequence, we assume that stars between $8-25\,{\rm M}_{\odot}$ undergo CCSNe, while more massive stars directly collapse into a black hole \citep{Janka2012, Zapartas+2021}. Less massive stars instead spend a relatively short period as giants. This phase lasts until their envelope has been ejected, and only the core remains as a white dwarf.

Massive ($\geq8\,{\rm M}_{\odot}$) stars eject fast ($\sim1000\,{\rm km\,s}^{-1}$) winds while on the main-sequence \citep{Vink+2015}. During this time, the mass loss rate is in the range $10^{-8}-10^{-6}\,{\rm M}_{\odot}\,{\rm yr}^{-1}$. In our model, the exact rate is determined by the total mass ejected during this phase, divided by the main-sequence lifetime. This is an approximation as winds from O and B-type stars ramp up significantly during the final stages of the main-sequence. Therefore, winds from individual stars are overestimated at early times with our constant mass-loss rate, but underestimated at the end.

Low mass stars inject slow ($10\,{\rm km\,s}^{-1}$) winds in pulses with mass loss rates $10^{-8}-10^{-4}\,{\rm M}_{\odot}\,{\rm yr}^{-1}$ \citep{Hofner&Olofsson2018}. In our model, we approximate these pulses by a constant rate $10^{-5}\,{\rm M}_{\odot}\,{\rm yr}^{-1}$; this rate determines the duration of the wind from the total mass ejected. As with massive stars, the amount of ejecta is determined by yields (Section~\ref{subsec:yield_tables}). With the yields applied in this work, stars match the initial-final mass relation for these types of stars \citep{Cummings+2016}.

To model the rate of SNeIa, we stochastically sample a discrete number of events from a delay-time distribution at each timestep $\Delta t$. The delay-time distribution is given by
\begin{equation}\label{eqn:Ia_model}
    n_{\rm Ia}=I_{\rm Ia}\, M_{\rm sf}\,\left[\frac{t}{\rm Gyr}\right]^{-1.12}\Delta t,\quad t>t_{\rm Ia},
\end{equation}
where $I_{\rm Ia}=2.6\times10^{-13}\,{\rm yr}^{-1}\,{\rm M}_{\odot}^{-1}$ is a normalization to the observed field SNIa rate \citep{Maoz&Graur2017}, and $M_{\rm sf}\lesssim500\,{\rm M_{\odot}}$ is the total mass used to sample stars at each star formation event. The delay-time $t_{\rm Ia}$ parametrizes the time after star formation before which any SNeIa can occur. Our fiducial choice is $t_{\rm Ia}=38\,{\rm Myr}$, motivated by the main sequence lifetime of an ONe white dwarf progenitor. We will vary this parameter later since the exact progenitors of SNeIa are uncertain and this timescale can vary across the range $\approx38-400\,{\rm Myr}$ \citep{Maoz&Graur2017}.

\subsubsection{Element-by-element yields}\label{subsec:yield_tables}

The total and element-by-element mass loss from stars at different stages of stellar evolution is determined through bilinear interpolation of tabulated yields. While the method allows for tracking of any element in the provided yield table, we limit this work to exploring C, N, O, Mg, Al, Si, and Fe. Briefly summarized: C and N are ejected from stars of different masses on the AGB (resulting in different timescales for enrichment); O, Mg, and Si are primarily produced in CCSN and trace chemical enrichment on $0.1-1\,{\rm Gyr}$ timescales when measured relative to Fe; Al production is sensitive to the amount of C in the core after He burning, and therefore depends on stellar evolution modeling. As in previous \textsc{edge} models, we define total metallicity from $M_{Z}=2.09\,M_{\rm O} + 1.06\,M_{\rm Fe}$ \citep{Agertz+2020}\footnote{To calculate abundances relative to hydrogen, we assume a primordial helium fraction ($Y=0.246$) and calculate the hydrogen fraction $X=(1-Z)(1-Y)$, i.e., assuming a constant Y at all $Z$. We leave to future work an explicit account of H and He enrichment.}. For stars with properties outside the tabulated ranges, we allow the method to extrapolate while ensuring that mass loss does not exceed $90\%$ of the star's initial mass. Tables are divided into different types of sources: mass loss from $0.5-8.0\,{\rm M}_{\odot}$ stars after main sequence (henceforth referred to as AGB ejecta); mass loss from $8.0-100\,{\rm M}_{\odot}$ mass stars during main sequence (henceforth referred to as winds); explosive ejection when $8.0-25\,{\rm M}_{\odot}$ undergo CCSNe; explosive ejection for SNeIa.

In this work, all simulations use the same model for AGB ejecta from \textsc{NuGrid} \citep{Pignatari+2016, Ritter+2018}, tabulated in 8 bins in stellar mass between $1.0-7.0\,{\rm M}_{\odot}$ and 5 bins in metallicities $0.005-1.0\,{\rm Z}_{\odot}$, where ${\rm Z}_{\odot}=0.02$. 

For winds and CCSN, we test two sets of yield tables from \textsc{NuGrid} \citep{Pignatari+2016, Ritter+2018} and three sets of yield tables from \citet{LimongiChieffi2018} (henceforth referred to as \citetalias{LimongiChieffi2018}). For \textsc{NuGrid}, the yield sets are distinguished by rapid and delayed explosion mechanisms for CCSN. The rapid explosion mechanism limits the time ($250\,{\rm ms}$) allowed for kinetic energy to be generated from neutrino pressure to overcome the kinetic energy of in-falling material \citep{Fryer+2012}. The delay allows for an explosion if this criterion is reached later. Each \textsc{NuGrid} model is tabulated in 4 bins in stellar mass between $12-25\,{\rm M}_{\odot}$ and with the same metallicity bins as for AGB ejecta. The yields from \citetalias{LimongiChieffi2018} provides tables for rotation velocities $0$, $150$ and $300\,{\rm km\,s}^{-1}$, tabulated in 4 metallicity bins between $0.0016 - 0.675\,{\rm Z}_{\odot}$ and in 9 bins between stellar masses $13-120\,{\rm M}_{\odot}$. In addition to testing each rotation model separately, we include a simulation where each star is assigned rotation at birth based on randomly sampling the rotation distributions from \citet{Prantzos+2018}. This rotation is then used to determine which of the yield tables from \citetalias{LimongiChieffi2018} to use on a star-by-star basis.

For SNeIa, we use yields from detonation models of Chandrasekhar-mass white dwarfs with equal initial mass in carbon and oxygen \citep{Seitenzahl+2013}. We use their model N100, which is provided in four different metallicities covering the range $0.0065-0.65\,{\rm Z}_{\odot}$. The metallicity dependence is weak, and in most of our models, we use only the highest metallicity bin for simplicity. We include one simulation where the yields are interpolated in metallicity.

%----------------------------
\subsubsection{Enrichment from a single stellar population}\label{subsec:yield_summary}
%----------------------------

\begin{figure}
    \centering
    \includegraphics[width=\linewidth]{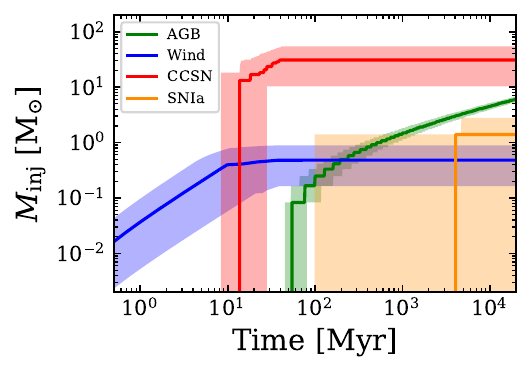}
    \caption{Median total mass loss as a function of time, with colored regions showing the $68$ percentiles for $1000$ samples of $500\,{\rm M_{\odot}}$ drawn from the IMF. This example uses the yields from \textsc{NuGrid}, and type Ia delay time of $38\,{\rm Myr}$, but the scatter in injected mass at a given time is similar regardless of model. Note that the large scatter in SNIa arises because of forcing discrete events for a small amount of stellar mass.}
    \label{fig:stochasticity}
\end{figure}

We show the cumulative mass ejected of tracked elements and total metallicity ($Z$) from the studied yields models in Figure~\ref{fig:yields}. In this example, each model is evolved for $500\,{\rm M}_{\odot}$ of stars that are formed instantaneously (i.e., the behavior of each star formation event, later described in Section~\ref{sec:numerical_method}). Except for Fe and, in some models, N, the majority of mass in each element, as well as total metallicity, originates from CCSNe when summed over $13.8\,{\rm Gyr}$, regardless of model. Fe production is dominated by SNIa, albeit only after $\sim1\,{\rm Gyr}$ of evolution.

Yields from winds are relatively low compared to CCSN; however, the enrichment starts right after star formation. This makes self-enrichment in star-forming regions and stellar clusters possible (local star formation is typically self-quenched on timescales $\lesssim10\,{\rm Myr}$, \citealt{Andersson+2024}). Comparing \textsc{NuGrid} and \citetalias{LimongiChieffi2018}, the latter has significantly stronger winds at high stellar rotation (at no rotation, C, N, and O yields are lower in \citetalias{LimongiChieffi2018} and comparable to \textsc{NuGrid} for heavier elements). 

Similarly, CCSN models with rotation significantly enhance the production of N, O, Al, and Si (and a small enhancement in C). Except for N (which is decreased) and Al (which is increased), 0 km/s rotation yields in \citetalias{LimongiChieffi2018} CCSN models are comparable to those in \textsc{NuGrid}. The total CCSN Fe production is similar between \citetalias{LimongiChieffi2018} and \textsc{NuGrid}; however, the evolution differs in the period when CCSNe are exploding ($8-40\,{\rm Myr}$). Very massive stars produce more Fe in all models from \citetalias{LimongiChieffi2018} (early rapid increase in Fe); however, \textsc{NuGrid} quickly catches up when stars around $8-12\,{\rm M}_{\odot}$ start exploding. This is partly due to extrapolation of the yield table (see Section~\ref{subsec:yield_tables} for further details about the extrapolation method), which only extends to $12\,{\rm M}_{\odot}$; therefore, the Fe yields in our implementation of the \textsc{NuGrid} yields are likely overestimated (see also discussion in \citealt{Rey+2025}).

Compared with other sources (except winds), AGB stars produce large amounts of C and N (note that compared to SNIa, $Z$ production is comparable, and O production is higher). Importantly, AGB stars behave similarly to SNIa, producing enriched material over several Gyr, particularly C and O.

SNIa are important sources of Si, Fe, and $Z$, although we note that O production is similar to that of AGB stars. While SNIa typically occur on longer timescales, each event produces a large amount of material. Therefore, on average, SNIa enrichment is reasonably high in terms of mass in the first few hundred $\rm Myr$s. Because of this, the assumption of $t_{\rm Ia}$ is important and plays a significant role in early enrichment of elements originating from SNIa. We exemplify the effect of sampling discrete SNeIa in Figure~\ref{fig:stochasticity},  showing the tracks representing the median and $68$th percentiles of a $500\,{\rm M}_{\odot}$ stellar population, assuming $t_{\rm Ia}=38\,{\rm Myr}$. On average, one such population produces $0.85$ SNeIa in a $13.8\,{\rm Gyr}$, while the same number for our model assuming $t_{\rm Ia}=100\,{\rm Myr}$ is $0.64$. For our simulations, the dwarf galaxies experience approximately $190$ and $130$ SNeIa for a $t_{\rm Ia}$ of $38\,{\rm Myr}$ and $100\,{\rm Myr}$, respectively. Furthermore,  Figure~\ref{fig:stochasticity} shows that only $50\%$ of $500\,{\rm M}_{\odot}$ populations produce a SNIa within the first $4\,{\rm Gyr}$; however, $16\%$ produce one within the first $\approx 100\,{\rm Myr}$. For comparison, in models with $t_{\rm Ia}=100\,{\rm Myr}$, $16\%$ of $500\,{\rm M}_{\odot}$ populations produce one within the first $\approx 300\,{\rm Myr}$.

All feedback (including the ejection of chemical elements) is determined on a star-by-star basis; therefore, feedback is inherently stochastic from sampling of the IMF and timing of events. This implies that each quanta of $500\,{\rm M}_{\odot}$ produces different amounts of ejecta, as discussed for SNeIa in the previous paragraph. Figure~\ref{fig:stochasticity} quantifies this scatter for other sources as well, showing the median and $68$th percentiles of the cumulative mass loss as a function of time for one of our models. Noise sampling of the IMF generates factor-of-a-few changes in injected mass, comparable in scale to some of the differences between yield models. This, combined with other sources of stochasticity in galaxy formation modeling, motivates the quantification of stochastic effects in Section~\ref{sec:result}. 

%----------------------------
\subsection{Numerical method \& initial conditions}\label{sec:numerical_method}
%----------------------------

\begin{table*}[]
    \centering
    \caption{First column gives the labels used throughout the paper, with simulations divided into four different groups based on the assumption that is varied. Note that some simulations appear multiple times in the table, and the final two labels represent three simulations each. When applicable, the second column gives stellar rotation, and the third column gives the delay time ($t_{\rm Ia}$, see Equation~\ref{eqn:Ia_model}) for SNIa. The last two columns include notable features of each model and a list of yield table references. Reference are abbreviated by number accordingly: [1] \citet{Pignatari+2016}; [2] \citet{Ritter+2018}; [3] \citet{Seitenzahl+2013}; [4] \citetalias{LimongiChieffi2018}; [5] \citet{Prantzos+2018}; [6] \citet{Fryer+2012}. For [3] marked with a star, we only apply yields from model N100 from \citet{Seitenzahl+2013}, i.e., SNIa yields at solar metallicity.}
    \begin{tabular}{l l l l l}
        \hline\hline
        \hline
        Label & Stellar rotation & $t_{\rm Ia}$ & Note & Yield table \\
        \hline\hline
         \textsc{NuGrid} (delay) & N/A & $38\,{\rm Myr}$ & Allows delayed trigger (CCSN), see [6] & [1,2,3*] \\        
         \textsc{NuGrid} (rapid) & N/A & $38\,{\rm Myr}$ & W/o delayed trigger (CCSN), see [6] & [1,2,3*] \\
         \hline
         LC18 (0 km/s) & $0\,{\rm km\,s}^{-1}$  & $38\,{\rm Myr}$ & & [3*,4] \\
         LC18 (150 km/s) & $150\,{\rm km\,s}^{-1}$  & $38\,{\rm Myr}$ & & [3*,4] \\
         LC18 (300 km/s) & $300\,{\rm km\,s}^{-1}$  & $38\,{\rm Myr}$ & & [3*,4] \\
         LC18 (Prantzos et al.) & $[0,150,300]\,{\rm km\,s}^{-1}$  & $38\,{\rm Myr}$ & Stochastically sampled rotation, see [5] & [3*,4] \\
         \hline
         $t_{\rm Ia}=38\,{\rm Myr}$ & $150\,{\rm km\,s}^{-1}$  & $38\,{\rm Myr}$ & Same as LC18 (150 km/s) & [3*,4]\\
         $t_{\rm Ia}=100\,{\rm Myr}$ & $150\,{\rm km\,s}^{-1}$  & $100\,{\rm Myr}$ & & [3*,4] \\
         w/o SNe Ia & $150\,{\rm km\,s}^{-1}$  & N/A & No Ia enrichment & [4] \\
         Z dep. yields & $150\,{\rm km\,s}^{-1}$  & $38\,{\rm Myr}$ & Metallicity dependent SNIa yields & [3,4]\\
         \hline
         \textsc{NuGrid} (delay), $t_{\rm Ia}=38\,{\rm Myr}$ & N/A & $38\,{\rm Myr}$ & Multiple runs of \textsc{NuGrid} (delay) & [1,2,3*] \\   
         LC18 (150 km/s), $t_{\rm Ia}=100\,{\rm Myr}$ & $150\,{\rm km\,s}^{-1}$  & $100\,{\rm Myr}$ & Multiple runs of LC18 (150 km/s) & [3*,4] \\
         \hline
    \end{tabular}
    \label{tab:models}
\end{table*}

Our numerical framework builds on a modified version of the adaptive mesh refinement (AMR) and $N$-body code \textsc{ramses}\footnote{https://ramses-organisation.readthedocs.io} code \citep{Teyssier2002}. The fluid equations are solved using a second-order unsplit Godunov method (HLLC) for the Riemann solver \citep{Toro+1994} with a MinMod slope limiter. Gravitational dynamics are treated via the Poisson equation using the multi-grid method \citep{Guillet&Teyssier2011}, treating all stellar and dark matter as collisionless particles with mapping to the AMR grid by the cloud-in-cell particle-mesh method. We track the mass fraction of elements by advecting passive scalars $f_i$ in time $t$ with the fluid flow following
\begin{equation}
    \frac{{\partial}(\rho f_{i})}{\partial t} = -\nabla(\rho f_i\,\mathbf{v}),
\end{equation}
where $\rho$ and $\mathbf{v}$ is the fluid density and velocity, respectively. 

Gas cooling and heating are described in detail in \citet{Rey+2020}. In summary, we assume an ideal monoatomic gas with adiabatic index $\gamma=5/3$, and limit temperatures to the range $1-10^9\,{\rm K}$. Our routine accounts for hydrogen and helium thermochemistry assuming equilibrium conditions \citep{Courty&Alimi2004, Rosdahl+2013} and metal line cooling based on \textsc{cloudy} \citep{Ferland+1998}. In addition, we add photo-ionization and photo-heating rates from a UV background modified from \citet{Haardt&Madau1996}, including a suppression of these rates in dense, self-shielded gas \citep{Aubert&Teyssier2010, Rosdahl&Blaizot2012}. To account for the unresolved formation of Population III stars in our simulations, we seed the zoom region with an oxygen fraction of $2\times10^{-5}$. Early star formation is insensitive to the exact choice of this floor \citep{Agertz+2020}, but the lack of metal-free star modeling will lead to spurious abundances at very low [X/Fe]. Out of caution, we remove stars with [Fe/H]$<-4$ in our analysis, as these might have abundances affected by enrichment from Pop.~III stars \citep{Jeon+2017}.

All simulations in this work start from the same initial conditions and employ a zoom-in refinement strategy to target the formation of a single galaxy in $\Lambda$ cold dark matter cosmology. Our initial condition are generated using \textsc{genetIC} \citep{Roth+2016, Rey&Pontzen2018, Stopyra+2021}. They assume cosmological parameters from \citet{Plank+2014} and include targeted changes to the growth history using the `genetic modification' method \citep[for more details, see][]{Rey+2019, Agertz+2020, Andersson+2025}. The zoom-in region is defined to include all particles within two times the virial radius at $z=0$, obtained beforehand from a $512^3$ dark matter only simulation \citep{Rey+2019, Rey+2020}. For simulations presented in this work, dark matter particles are given a mass of $940\,{\rm M}_{\odot}$ within the zoom-in region, and step-wise increased by factors of 8 up to a mass of $2.5\times10^{8}\,{\rm M}_{\odot}$, using a buffer region of at least 8 grids width between each step\footnote{Note that the multi-grid method used to calculate gravitational forces is well-suited to deal with the disparity between particle masses, which is important for systems with massive dark matter particles and low mass stars, which in our case can be $940\,{\rm M}_{\odot}$ dark matter particles interacting with $0.5\,{\rm M}_{\odot}$ star particles.}. Furthermore, the AMR grid is adapted from 7 to 25 levels of refinement (maximum resolution of $3.6\,{\rm pc}$ at $z=0$) to keep the cell baryon (stars+gas) mass around $167\,{\rm M}_{\odot}$. Furthermore, refinement is set to maintain approximately 8 dark matter particles per cell, which reduces discreteness effects \citep{Romeo+2008}.

%----------------------------
\subsubsection{Star formation and feedback}\label{subsec:inferno}
%----------------------------

Star formation, stellar evolution, and feedback are treated by the \textsc{inferno} model described in detail in \citet{Andersson+2023}. A star formation check is triggered when gas cells within the refinement region exceed a density of $\rho=300\,{\rm cm}^{-3}$ if the gas is colder than $1000\,{\rm K}$. In cells satisfying this criterion, we compute a star formation rate density
\begin{equation}
    \dot{\rho}_{\star}=\epsilon_{\rm ff}\frac{\rho}{t_{\rm ff}},
\end{equation}
where $t_{\rm ff}$ is the free-fall time of the gas cell, and $\epsilon_{\rm ff}=0.1$ is an assumed star formation efficiency per free-fall time. The parametrization of our star formation recipe has been shown to successfully match key observables, both in \textsc{edge} \citep[see, e.g.,][]{Rey+2019, Agertz+2020, Rey+2025}, and \textsc{inferno} \citep{Andersson+2020, Andersson+2023, Andersson+2025} simulations. Based on the $\dot{\rho}_{\star}$, star formation is actualized in discrete quanta of $M_{\rm sf}=500\,{\rm M}_{\odot}$ using Poisson sampling.

At star formation, individual stars are sampled from the IMF by \citet{Kroupa2001}, given by
\begin{equation}\label{eq:imf}
    \xi(m) = AC_im^{-a_i},
\end{equation}
with two intervals denoted with $i$ between $0.08-100\,{\rm M}_{\odot}$ split at $m_0=0.5\,{\rm M}_{\odot}$. In Equation~\ref{eq:imf}, $A$ is a normalizing constant, $C_0=1.0$ and $C_1=C_0m_0^{a_1-a_0}$ are constants ensuring continuity, $a_0=1.3$ below $m_0$ and $a_1=2.3$ above. To sample the IMF, the method from \citet{Sormani+2017} is used, with the same implementation as in \citet{Andersson+2020}. Stars with mass below $0.5\,{\rm M}_{\odot}$ are agglomerated into a single particle, while more massive stars are spawned as individual particles.

All stars inherit the velocity of the gas cell from which they formed. In addition, individual stars are given small velocity perturbations from a normal distribution with standard deviation of $0.01\,{\rm km\,s}^{-1}$. These small perturbations are not to be confused with natal kicks resulting in walk- and runaway stars previously used with \textsc{inferno} \citep{Andersson+2020, Andersson+2021, Andersson+2023}. To limit the uncertainty related to assumptions about fast-moving stars, we leave a study of these types of objects for future work.

As detailed in Section~\ref{sec:yield_method}, stellar evolution sets the time when stars inject mass, momentum, energy, and the chemical elements traced by scalar fields. All these quantities are injected into the nearest 8 cells following the implementation of \citet{Agertz+2013}. When active, winds are injected continuously at constant velocity; the wind velocity during AGB is $10\,{\rm km\,s}^{-1}$, and $1000\,{\rm km\,s}^{-1}$ for massive ($>8\,{\rm M}_{\odot}$) main sequence stars. CCSNe locations are determined by individual stars; however, because we do not track binary evolution, SNIa explosions have positions traced by the unresolved stellar populations. For these explosions, stars inject $10^{51}\,{\rm erg}$ of thermal energy and the corresponding momentum given the total mass ejected. For each explosion, we check if the cooling radius is resolved by at least 6 cells. If this is not the case, the scheme injects the terminal momentum based on the Sedov-Taylor solution calibrated by \citet{Kim&Ostriker2015}. Our implementation has been shown to reproduce the total momentum and energy injection even at low resolution \citep{Ohlin+2019}, and we note that $<10\%$ of explosions are in the unresolved regime \citep[see further discussion in][]{Rey+2025}.

Further description of the \textsc{inferno} model is provided in \citet{Andersson+2023}, including a presentation of the total mass, momentum, and energy budget. Furthermore, \citet{Andersson+2023} found convergence for star formation and feedback at the resolution in this work, given our choices of parameters.

%----------------------------
\subsection{Suite of simulations}\label{subsec:simulations}
%----------------------------

\begin{figure}
    \centering
    \includegraphics[width=\linewidth]{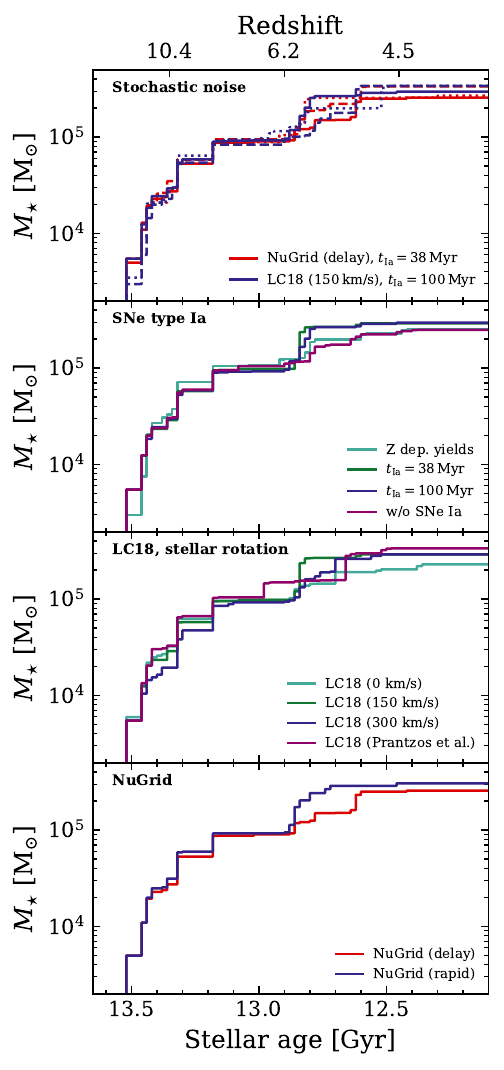}
    \caption{Cumulative stellar mass as a function of stellar age (bottom axis) and redshift (top axis). Simulations are divided into the same groups as in Table~\ref{tab:models}, emphasizing the assumption that is varied in each panel (each panel is labeled in the top left corner). The top panel shows three simulations using different lines but the same color for the two different models.}
    \label{fig:sfh}
\end{figure}

\begin{figure*}
    \centering
    \includegraphics[width=\linewidth]{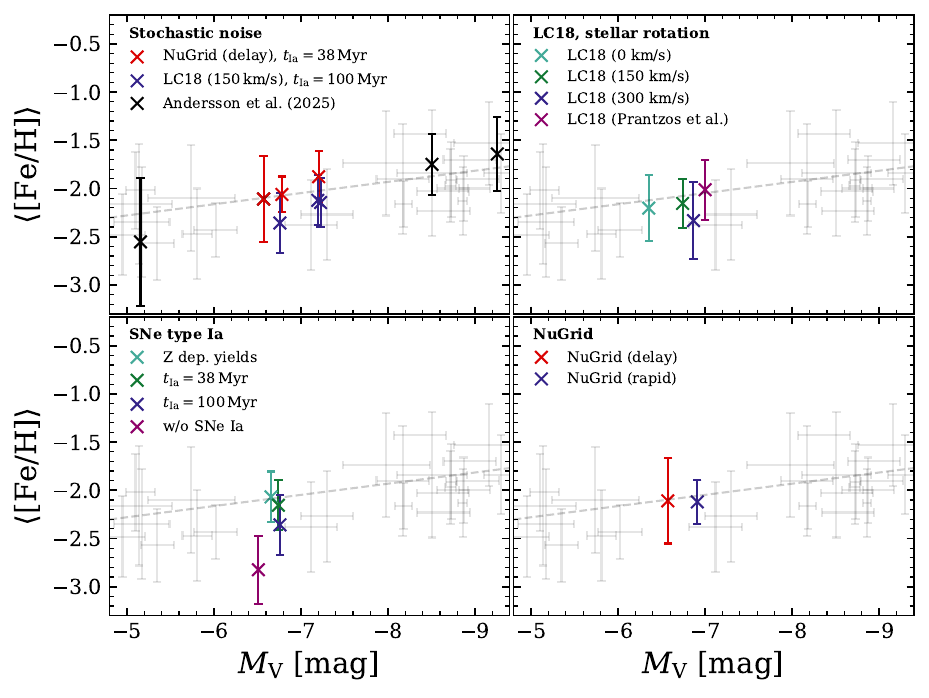}
    \caption{Mean iron abundance $\langle[{\rm Fe/H}]\rangle$ as a function of total V-band magnitude $M_{\rm V}$ for all simulations presented in this work, divided into four groups (see Table~\ref{tab:models}) for easier comparison. \textit{Top left}: exploring the role of stochasticity by fixing the subgrid model to be \textsc{NuGrid} (delay) with $t_{\rm Ia}=38\,{\rm Myr}$ (red) or \citetalias{LimongiChieffi2018} with $t_{\rm Ia}=100\,{\rm Myr}$ (blue) varying only the random number seed.  This panel also include the simulations from \citet{Andersson+2025} in black color. \textit{Top right}: exploring the rotation models from \citetalias{LimongiChieffi2018} (teal, green, blue for $0$, $150$, and $300\,{\rm km\,s}^{-1}$, respectively), including one example with the rotation distribution from \citet{Prantzos+2018} (purple). All these models use $t_{\rm Ia}=38\,{\rm Myr}$. \textit{Lower left}: testing different SNIa models, including $38$ (green) and $100\,{\rm Myr}$ (blue) for $t_{\rm Ia}$, one example without SNIa (purple), and one with metallicity-dependent SNIa yields (teal). All models use yields from \citetalias{LimongiChieffi2018} for massive stars. \textit{Bottom right}: comparing the \textsc{NuGrid} models with the delay (red) and rapid (blue) explosion triggers for CCSN, both with $t_{\rm Ia}=38\,{\rm Myr}$. These simulations use a model identical to \textsc{NuGrid} (delay), $t_{\rm Ia}=38\,{\rm Myr}$, but from different initial conditions. The error bars on the vertical axis indicate the dispersion in [Fe/H], calculated by taking the standard deviation of all stars in each galaxy. Gray error bars show values from dwarf galaxies in the Local Group Volume, taken from the database by \citet{Pace2024}. For these points, $M_{\rm V}$ has error bars indicating upper and lower errors, while error bars on [Fe/H] indicate the dispersion. The gray dashed line shows the dwarf galaxy luminosity-[Fe/H] fit from \citet{Kirby+2013}.}
    \label{fig:FeH_MV}
\end{figure*}

Our simulations are executed from the same initial conditions, applying different choices relevant to chemical evolution (see Figure~\ref{fig:yields}). The final halo mass (redshift $z=0$) of this galaxy is $M_{200}=1.4\times10^{9}\,{\rm M}_{\odot}$, with $M_{200}$ defined as the mass enclosed within the spherical radius $r_{200}=24\,{\rm kpc}$, where the density is 200 times the cosmic critical density. 

We then re-simulate this same object, changing the chemical enrichment parameters as summarized in Table~\ref{tab:models}. To summarize, we divide our experiments into different families: (1) two massive star yield sets from \textsc{NuGrid}; (2) four massive star yield sets from \citetalias{LimongiChieffi2018} with different stellar rotation; (3) variations in SNIa delay times and yields; and (4) results from repeated simulations with the same chemical assumptions but for which the random seed for galaxy and star formation has been varied\footnote{Our simulation code does not conserve the order of distributed arithmetic operations over time, so we simply re-run the same model on different core numbers to change the random seed}. 

Comparing (1) and (2) allows us to compare two state-of-the-art yields for massive stars available in the literature. (3) allows us to test the importance of SNIa and control the amount of iron without drastically affecting the abundances of light elements. Finally, galaxy formation and stellar abundances have several sources of intrinsic noise associated with them, from sampling the IMF in low-mass systems (e.g., Fig~\ref{fig:stochasticity}) to the general chaoticity of galaxy formation \citep[see, e.g.,][]{Genel+2019, Keller+2019, Pakmor+2025}. (4) allows us to quantify the intrinsic scatter in our predictions within a given model and provide a metric by which to establish the significance of observed shifts between models.  

We show the cumulative stellar mass of all simulations in Figure~\ref{fig:sfh}. As expected, different models lead to slightly different star formation histories and stellar mass assemblies. However, differences between chemical enrichment variations are consistent with differences induced by pure stochasticity within a single model (e.g., top panel, blue lines). As a result, we conclude that the stellar mass assembly is consistent across all chemical enrichment models. 

\section{Results}\label{sec:result}
% ------------------------------------
\subsection{The luminosity-[Fe/H] relation}\label{sec:mzr_mdf}
% ------------------------------------

\begin{figure}
    \centering
    \includegraphics[width=\linewidth]{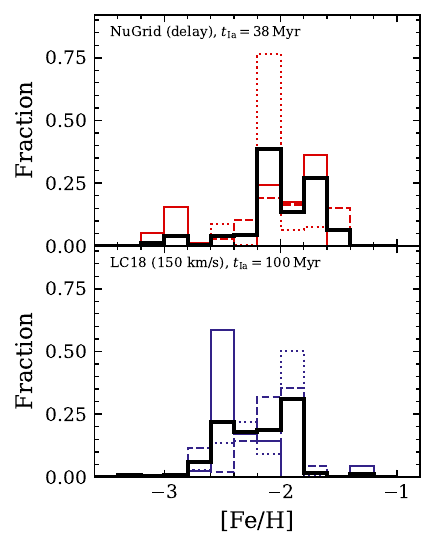}
    \caption{MDF of [Fe/H] of 6 simulations used to estimate the stochastic variance, each displayed showing the fraction of stars in bins with size 0.2 dex, comparable to the typical errors on observed estimates of these quantities. The colored lines show individual simulations, each distinguished by different lines, while the black lines show the combined MDF of all simulations using the same model. The top row shows results from the \textsc{NuGrid} (delay) model with SNIa delay time of $38\,{\rm Myr}$, and the bottom row shows results from the \citetalias{LimongiChieffi2018} model with $150\,{\rm km\,s}^{-1}$ rotation and a delay time of $100\,{\rm Myr}$ for SNIa. Each row includes three simulations (different line styles) executed with a different seed for random number sampling.}
    \label{fig:mdf_scatter}
\end{figure}

\begin{figure}
    \centering
    \includegraphics[width=\linewidth]{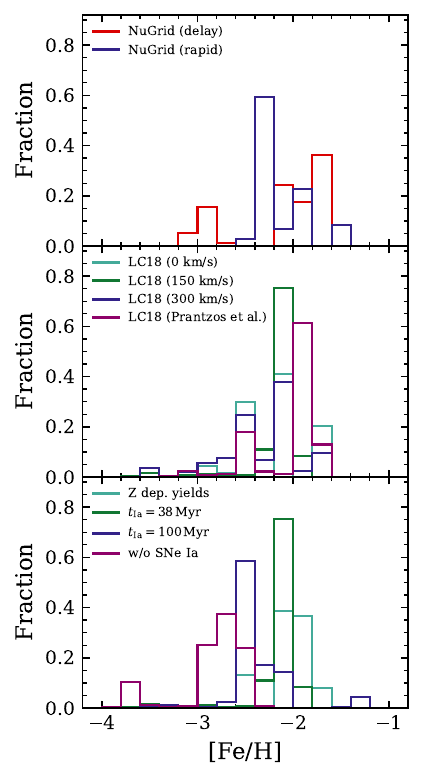}
    \caption{MDF of [Fe/H] (as in Figure~\ref{fig:mdf_scatter}) for the two models from \textsc{NuGrid} (top row), the four models including rotation (middle row), and the four different SNIa models (bottom row). See main text for details about each model.}
    \label{fig:mdf_compilation}
\end{figure}

\begin{figure*}
    \centering
    \includegraphics[width=\linewidth]{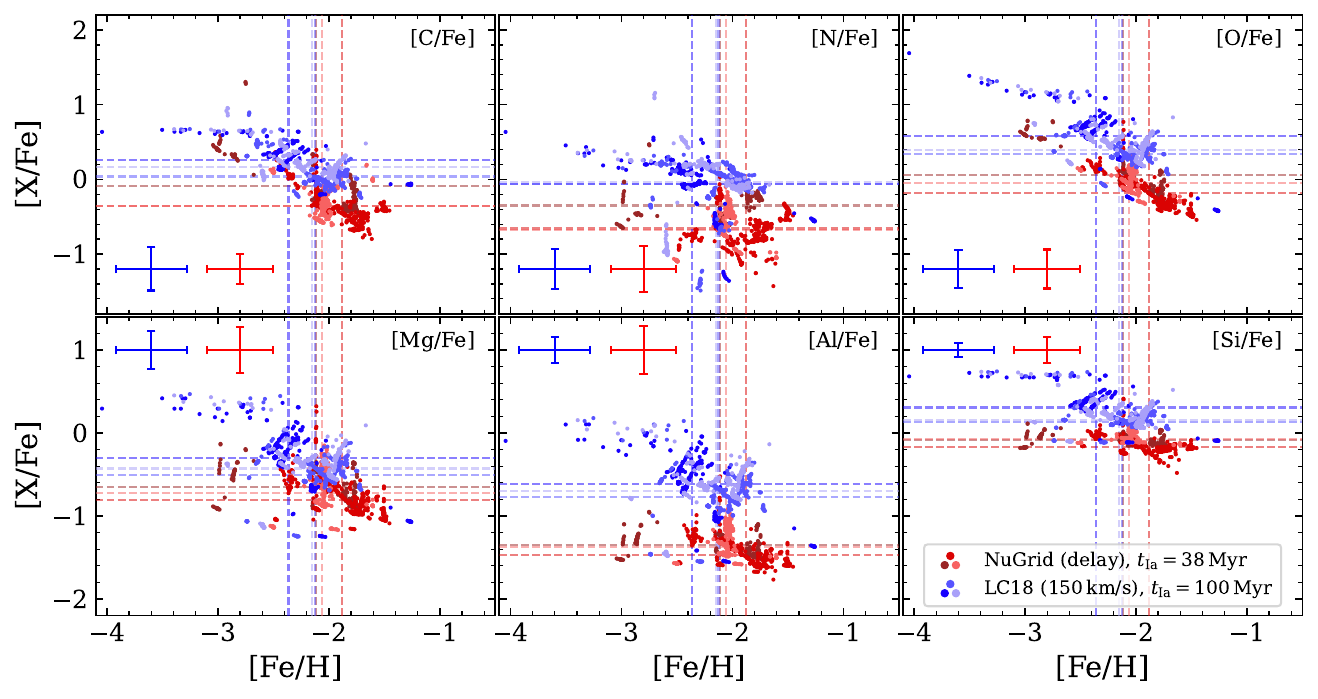}
    \caption{Abundances relative to Fe as a function of [Fe/H] for different elements as indicated by the label in the upper right corner of each panel. The figure includes data from two different models in different shades of red and blue, as indicated by the legend in the bottom right panel. For each model, different color shades present different simulations, varying only the random seed for IMF and SNIa sampling. The dashed vertical lines mark the mean value on the corresponding axis for each simulation. Included in the bottom (top) left of the top (bottom) row are colored error bars that show the total dispersion accounting for all stars in the three simulations of the corresponding color.}
    \label{fig:abundance_scatter}
\end{figure*}

Figure~\ref{fig:FeH_MV} shows the mean value of [Fe/H] as a function of total V-band magnitude for all simulations presented in this work. They are contrasted to estimates from observed dwarf galaxies in the Local Volume marked by gray error bars. Data points for observed systems show $M_{\rm V}$ and the associated upper and lower errors, while the [Fe/H] shows the spectroscopically determined values with errors indicating its dispersion. Note that vertical errors on the simulation data points also indicate [Fe/H] dispersion, not measurement error. The data was obtained from the database by \citet{Pace2024} and includes data from \citet{Carlin+2009, Kirby+2013, Martin+2014, Kirby+2015, Slater+2015, Li+2017, Simon2019, Pace+2020, Wojno+2020, Kirby+2020, Collins+2021, Jenkins+2021, Ji+2021, Charles+2023, Heiger+2024, Kvasova+2024, Muller+2025}.

Starting from the top-left panel, simulations with the same model but different random seeds evolve along the observed trend \citep{Kirby+2013}, but retain a clear ordering in which the \textsc{NuGrid} (delay) model with $t_{\rm Ia}=38\,{\rm Myr}$ always has higher $\langle[\rm Fe/H]\rangle$. 

We can disentangle the origin of this difference by comparing variations in massive star yields (right column) and in SNIa assumptions (bottom left). Considering only the position in $M_{\rm V}$ and mean [Fe/H], we find that different massive star yield tables cannot be distinguished from stochastic variations. The only potential exception is the extreme \citetalias{LimongiChieffi2018} model, for which all stars are assumed to rotate at 300 km/s, which shows a mean [Fe/H] 0.3 dex below the relation (blue, upper right panel). But accounting for the full rotation distribution increases the average [Fe/H] closer to the average relation (dashed line). By contrast, SNeIa have a much more noticeable impact, where delaying and eventually removing SNIa results in a [Fe/H] which is 0.6 dex below the \citet{Kirby+2013} relation. The systematic offset between stochastic re-simulations in our two models is thus readily attributed to their difference in SNeIa delay time distribution.

Overall, we conclude that the position in $M_{\rm V}$ and mean [Fe/H] is robust to the chemical enrichment assumptions tested, with our only simulation that can be ruled out being the model with no SNeIa at all. Our results also show that stochastic effects within a single or a couple of faint dwarf galaxies will make it extremely challenging to differentiate between chemical models based on position in $M_{\rm V}$ and mean [Fe/H] alone. To emphasize this point, we show the three different growth histories from \citet{Andersson+2025} all with the \textsc{NuGrid} (delay) model with $t_{\rm Ia}=38\,{\rm Myr}$ (black markers in top-left panel). Differences in mass assembly, star formation, and chemical enrichment histories at fixed halo mass lead to larger shifts in average [Fe/H] ($\approx 0.4$ dex) than intrinsic stochasticity ($\approx 0.2$ dex), but comparable to SNeIa variations. 

Nonetheless, chemical assumptions are likely to introduce systematic effects across the whole dwarf galaxy population and could be teased out through statistical modeling of populations (see e.g. \citealt{Rey+2025} for a visualization of this effect). Alternatively, a careful comparison between points in Figure~\ref{fig:FeH_MV} reveals that the spread in [Fe/H] (vertical errors) is somewhat sensitive to the choice of yields, highlighting that more detailed chemical observables (e.g., MDFs) might provide additional sensitivity to chemical modeling. We focus on this in the next Section. 

% ------------------------------------
\subsection{Metal distribution functions}\label{subsec:mdf}
% ------------------------------------

Figures~\ref{fig:mdf_scatter} and \ref{fig:mdf_compilation} show MDFs quantified by [Fe/H] for our simulations. The bin width of $0.2\,{\rm dex}$ is comparable to the typical observational error. In addition, we show the combined MDF of all three simulations in both panels with black lines. The only robust difference is the systematic shift towards lower [Fe/H] when increasing $t_{\rm Ia}$, as is already discussed in the previous Section.

For completeness, we show MDFs of the different models in Figure~\ref{fig:mdf_compilation}, dividing the groups outlined in Table~\ref{tab:models} between the different panels. Overall, the MDFs are similar across the different model assumptions. Changing the massive-star yields or the stellar rotation rate produces only minor differences in the distributions. In some cases, the number and overall shape of peaks vary slightly, but these effects remain well within the stochastic variance. The only systematic change that is apparent, given typical measurement uncertainties, is the aforementioned shift toward lower [Fe/H] when delaying or removing SNeIa. 

Note that we checked MDFs of other elements, including combinations of light elements ([$\alpha$/Fe]), reaching similar conclusions. Instead, we turn to trends between relative abundances in the next Section.

%---------------------------------------------
\subsection{Relative abundances of individual elements}\label{sec:relative_abundances}
%---------------------------------------------

\begin{figure*}
    \centering
    \includegraphics[width=\linewidth]{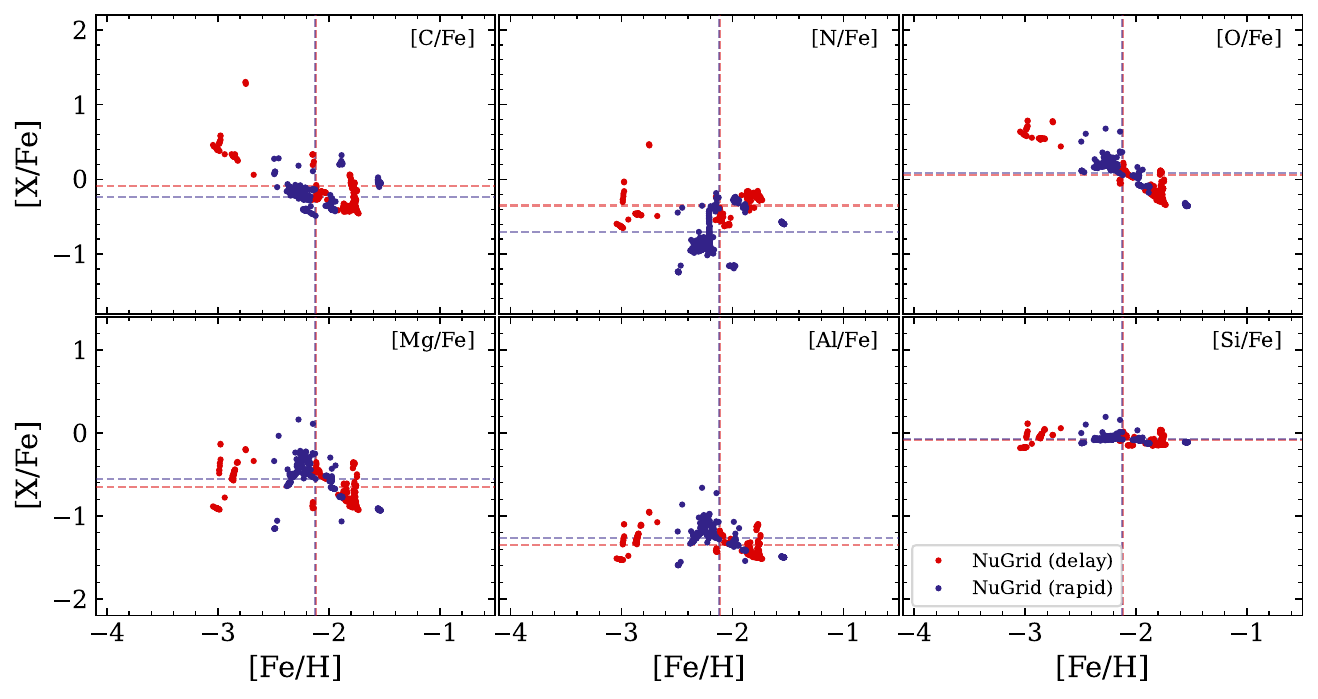}
    \caption{Abundances (relative to iron) for the elements in our simulations (indicated in the top right of each panel), all as a function of iron abundance (relative hydrogen). The figure shows results from simulations that apply yields from \textsc{NuGrid}, including the delayed and rapid timing models for core-collapse explosion trigger (see main text for details). The mean value of all stars along each axis is shown by dashed lines in the color assigned to the model (see legend).}
    \label{fig:abundance_nugrid}
\end{figure*}

\begin{figure*}
    \centering
    \includegraphics[width=\linewidth]{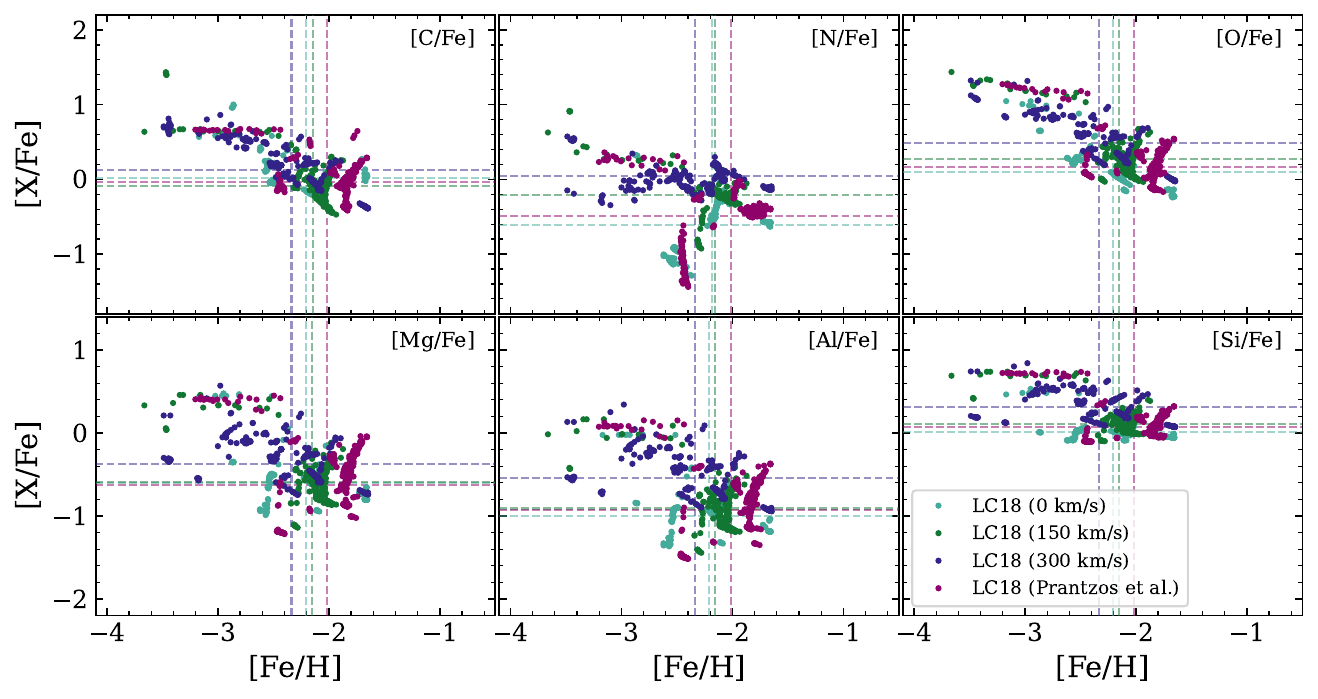}
    \caption{Same as Figure~\ref{fig:abundance_nugrid}, instead showing models with yields for massive stars from \citetalias{LimongiChieffi2018} with different stellar rotation (see legend). In addition to single rotation models, we include a model sampling the rotation of each star at birth using the distribution from \citet{Prantzos+2018} and selects the yield table accordingly.}
    \label{fig:abundance_rotation}
\end{figure*}

\begin{figure*}
    \centering
    \includegraphics[width=\linewidth]{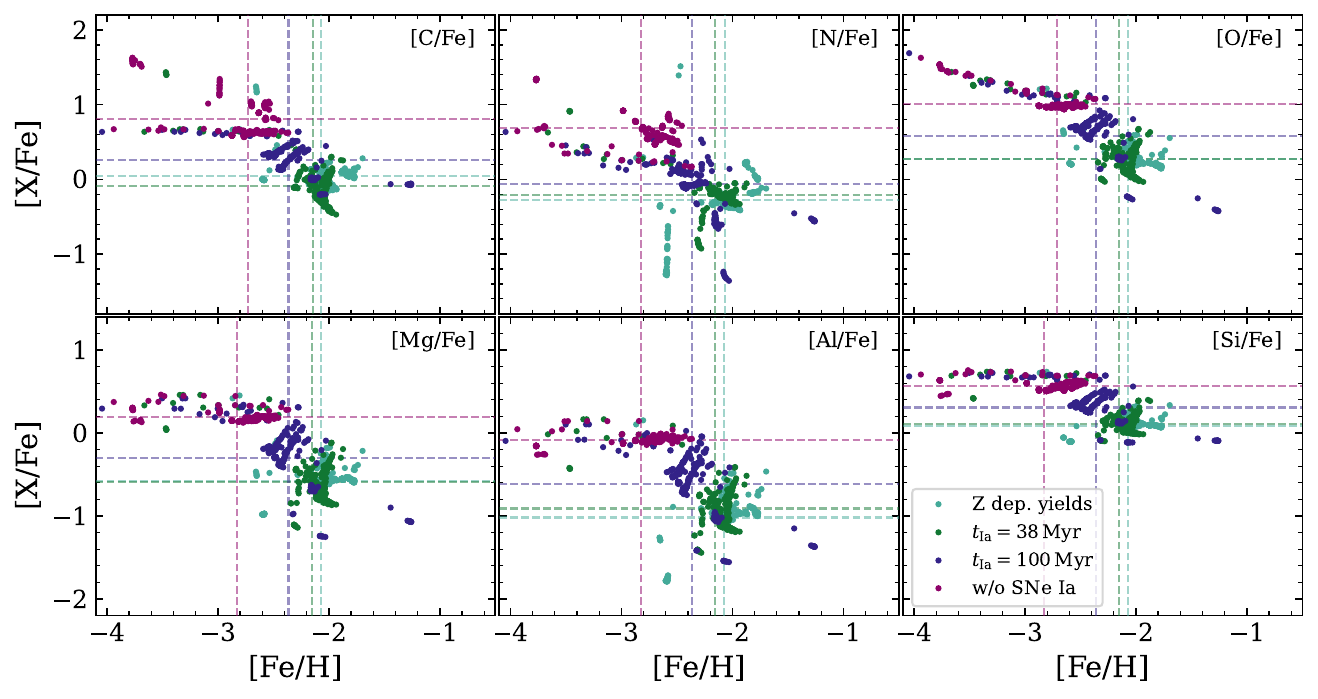}
    \caption{Same as Figure~\ref{fig:abundance_nugrid} for models with different assumptions for SNIa, including two different time delays ($t_{\rm Ia}=38\,{\rm Myr}$ and $t_{\rm Ia}=100\,{\rm Myr}$). In addition, we include one model without SNIa and one model where yields (taken from \citealt{Seitenzahl+2013}) are interpolated in metallicity (denoted Z dep. yields), but otherwise the same as the model with $t_{\rm Ia}=38\,{\rm Myr}$. All models take yields from \citetalias{LimongiChieffi2018} with $150\,{\rm km\,s}^{-1}$ stellar rotation for massive stars.}
    \label{fig:abundance_snia}
\end{figure*}

We turn to the abundances of individual elements, all presented relative to Fe in this work (i.e., [X/Fe] for element X) and shown as a function of [Fe/H]. Figure~\ref{fig:abundance_scatter} shows the abundances of our simulations exploring random stochasticity within two models; \textsc{NuGrid} (delay), $t_{\rm Ia}=38\,{\rm Myr}$ in red and \citetalias{LimongiChieffi2018} (150 km/s), $t_{\rm Ia}=100\,{\rm Myr}$ in blue. The different simulations can be distinguished by the difference in color shade. For all panels, we indicate the mean abundance of each simulation using a dashed line in the corresponding color. 

In all cases, we observe the expected anti-correlation between [X/Fe] and [Fe/H]. As already noted when studying Figure~\ref{fig:FeH_MV}, the mean [Fe/H] is higher for the \textsc{NuGrid}, $t_{\rm Ia}=38\,{\rm Myr}$ runs due to the faster onset of SNeIa. As expected, this also correlates with systematically higher average [X/Fe] for all elements. 

Furthermore, the star-by-star scatter obtained when combining all simulations with the same model is shown by errors in the bottom left of each panel. Note that the scatter averaged over a wide range of [Fe/H] does not necessarily quantify the variance at a given [Fe/H] \citep{Mead+2024}, rather measuring the width of each element metal distribution function (note the limitation that the total distribution is typically multi-peak). Comparing the two models, we note that [C/Fe] evolves more (larger scatter) in \textsc{NuGrid} (delay), $t_{\rm Ia}=38\,{\rm Myr}$, while [Al/Fe] and [Si/Fe] evolve less (lower scatter). The other elements show similar scatter in the two models. Except for [Al/Fe] in \textsc{NuGrid} (delay) and [Si/Fe] in both models, the star-by-star scatter is typically around 0.6 dex. 

In addition to these broad trends in average abundances and scatter, element-by-element features appear that are related to the choice of yields. We will discuss these in the next sections, but we note that outliers exist in almost all simulations, either with very high or very low [X/Fe] compared to their simulation average. We have tied these individual outlier stars to specific enrichment events, on which we will report in future work. 

\subsection{Abundance trends from different assumptions}\label{subsec:abundance_trends}

Figures~\ref{fig:abundance_nugrid}, \ref{fig:abundance_rotation}, and~\ref{fig:abundance_snia} show the chemical abundances of individual stars in our three families of chemical assumptions. 

Focusing first on Figure~\ref{fig:abundance_nugrid}, we find that the two sets of \textsc{NuGrid} yields give highly consistent results for most elements that cannot be distinguished from differences introduced by stochasticity. As an example, the largest difference occurs in the mean [N/Fe] ($\approx - 0.35$ dex), which is comparable to stochastic shifts for [N/Fe] within the \textsc{NuGrid} model (Figure~\ref{fig:abundance_scatter}, top center). 

Similarly, Figure~\ref{fig:abundance_rotation} shows that the tables from \citetalias{LimongiChieffi2018} are mostly consistent with each other and compared to intrinsic stochasticity. The largest shifts are observed for the extreme assumption that all massive stars rotate with velocities of $300\,{\rm km\,s}^{-1}$, driving an over-production of N, O, Mg, Al, and Si. This over-production is a direct result of the stellar evolution at increasing velocity (Figure~\ref{fig:yields}). For example, Mg and Al production is increased due to the decreased amount of C left after the He burning phase in the core as rotation velocities increase \citep{LimongiChieffi2018}. Similarly, increasing stellar rotational velocity significantly augments the N output through early massive star winds (Figure~\ref{fig:yields}; see also \citealt{Meynet+2002, Chieffi&Limongi2013}). 

Again, we find much more significant differences in the mean abundance ratios when varying assumptions related to SNeIa (Figure~\ref{fig:abundance_snia}). Delaying SNeIa from 38 to 100 Myr drives all element ratios up by up to $\geq 0.4$ dex. Removing SNeIa entirely is even more dramatic, generating shifts up to $\approx 1$ dex, much larger than stochastic effects. These increases are entirely driven by reducing the iron content across the board, rather than increasing the production of specific elements. These results emphasize the key role played by SNeIa in setting the chemistry of UFDs, despite their short star formation histories ($\leq 2$ Gyr).

Now focusing on broader trends beyond the mean and comparing families of simulations, we see markedly different behavior in [X/Fe]-[Fe/H] between \textsc{NuGrid} and \citetalias{LimongiChieffi2018}, most clearly visible in [C/Fe], [O/Fe], and [Si/Fe]. Comparing Figures~\ref{fig:abundance_nugrid} and \ref{fig:abundance_rotation}, we find that \textsc{NuGrid} models show a single relation between these elements and [Fe/H], while \citetalias{LimongiChieffi2018} models appear bi-modal with a knee around [Fe/H]$=-2.5$, particularly in [Mg/Fe] and [Al/Fe]. This difference appears despite the same value for $t_{\rm Ia}$ and is robust compared to stochastic variance (see Figure~\ref{fig:abundance_scatter}). 

This knee is a well-established property of larger galaxies like the Milky Way \citep{Bensby+2014, Bland-Hawthorn+2019, Feuillet+2019} and is usually interpreted as the point in the star formation history where SNeIa start contributing significantly to iron production. Our results corroborate these ideas, with position for this knee around [Fe/H]$=-2.5$ not varying significantly with our assumptions for $t_{\rm Ia}$, but the feature (and any stars with [Fe/H]$>-2.5$) disappearing entirely when SNeIa are removed (Figure~\ref{fig:abundance_snia}). Note that the relative insensitivity of the knee position to the Ia onset timing is to be expected, as the variations implemented in this work are small compared to the overall length of the star formation histories ($\approx$ 1-2 Gyr). 

A notable feature in the simulations without SNeIa (purple in Figure~\ref{fig:abundance_snia}) is the presence of two distinct trends at low [Fe/H] in both [C/Fe] and, to a lesser extent, [N/Fe]. At lower [C/Fe], the abundance ratios are set by the yields of massive stars, whereas the rising trend originates from the accumulation of AGB material in small halos prior to accretion onto the main progenitor. These halos also remain quiescent for extended periods between star formation. An AGB-driven formation for carbon-enhanced metal-poor stars has been recently suggested by \citet{Gil-Pons+2025} and would require careful unpicking since this population is often linked with pollution from the first stars \citep{Klessen&Glover2023}. We will provide a more detailed analysis of this feature in a forthcoming study (Andersson et al., in prep.).

%---------------------------------------------
\section{Other sources of uncertainties yet to be tested}\label{sec:discussion}
 
In this work, we test 13 different models of chemical enrichment with a high-resolution, star-by-star cosmological zoom simulation. Despite this extensive suite, the computational cost and limited complexity of \textsc{inferno} restricted the breadth of the parameter space exploration. In this section, we highlight key areas that were not explored and would warrant further attention by future studies.

\subsection{Additional sources of enrichment and the limitations of tabulated yields}\label{subsec:other_yields}

We included variations of two sets of stellar feedback yields for massive stars, \textsc{NuGrid} \citep{Pignatari+2016, Ritter+2018} and rotation-dependent yields from \citetalias{LimongiChieffi2018}. For low mass stars, we tested only yields from \textsc{NuGrid}, and for SNIa we applied yields from \citet{Seitenzahl+2013}. 
As discussed in Section~\ref{sec:introduction}, this set of models is limited compared to the number of tabulated yields available in the literature.

Nonetheless, \citet{Buck+2021} systematically tested a broader range of yields in higher-mass systems up to Milky-Way mass, concluding that the total metallicity and iron content are broadly insensitive to choice of massive star yields, but that individual elements and trends in [X/Fe] - [Fe/H] are. Our conclusions extend these findings to very low-mass systems, giving confidence that our conclusions remain robust even if a wider exploration of massive star yield models were to be considered. 

Furthermore, the sources of chemical enrichment remain incomplete in our models. The most pressing limitation is the missing chemical yields for intermediate stellar masses. For low mass stars ($<8\,{\rm M}_{\odot}$), we applied the tables from \citep{Pignatari+2016, Ritter+2018}, with data points between $1$ and $7\,{\rm M}_{\odot}$. For higher mass stars, these tables cover the mass range $12$ and $25\,{\rm M}_{\odot}$, while the tables from \citepalias{LimongiChieffi2018} are in the range $13$ to $120\,{\rm M}_{\odot}$. This leaves a large range for extrapolation, a known issue for our simulations \citep{Rey+2025}. In fact, for the IMF applied here, this results in extrapolated yields for up to $\approx50\%$ (in terms of number) of all massive stars.

Yields for stars in the intermediate mass range ($7-12\,{\rm M}_{\odot}$) are particularly challenging to derive due to their complex evolution \citep{Doherty+2015, Doherty+2017, Gil-Pons+2018, Limongi+2024}. At the upper end of this mass range, these stars can result in electron-capture SNe, while stars at the lower end might result in exotic SNe \citep{Foley+2013}. However, the exact mass range for electron-capture SNe is sensitive to metallicity and remains uncertain, although it is likely narrow \citep{Doherty+2015, Gil-Pons+2018}. Nonetheless, chemical yields in this mass range are available \citep[e.g.,][]{Siess2010, Doherty+2014a, Doherty+2014b, Gil-Pons+2022, Limongi+2025} and we are currently working on adding these to our model. In doing so, we will address the effects of altering the boundary where the final stage of a star is a white dwarf.

Many other intriguing questions remain, in particular for the chemical signature in ultra-faint dwarf galaxies. For example, because their low mass places them at the limit where metal retention is possible, dwarf galaxies can place constraints on the production of r-process elements, believed to originate from rare events such as exotic forms of SNe (collapsar and magnetorotational) and neutron-star mergers \citet{Ji+2016, Cote+2019, vandeVoort+2020, Tarumi+2020}. Furthermore, significant contributions from exotic forms of SNIa (e.g., Iax from sub-Chandrasekhar mass primogenitors) might be necessary to fully explain dwarf galaxy abundances \citet{Kobayashi+2020b}. Furthermore, a complete picture of chemical abundances most likely demands treatment of binary stars \citep{Nguyen+2024}, including common envelope overflow and novae, even in dwarf galaxies \citep{Yates+2024}. These are some of the many sources that are yet to be fully explored with hydrodynamical models, providing promising venues for future work. We refer to \citet{Kobayashi+2020} for further discussion about additional sites of nucleosynthesis.

\subsection{Zero-metallicity stars}\label{subsec:popIII}

In this paper, we do not consider the formation of stars from primordial gas; instead, we impose a homogeneous oxygen floor to represent this unresolved phase of chemical enrichment. In Appendix~\ref{sec:metal_floor}, we show that this floor has a minimal contribution to all oxygen abundances presented in this work. 

Here, we also justify that the simplicity of this assumption is justified at the mass scale and environment we consider. Pop.~III stars have been shown to make significant contributions to the mean iron content of UFDs when simulated in a proto Milky-Way environment (e.g. \citealt{Rey+2025b}), but much less in isolated field UFDs like we study here \citet{Jeon+2017, Sanati+2023}. Furthermore, the chemical contribution from Pop.~III stars is increasingly washed out as stellar masses increase towards the `massive' end of UFDs ($10^5\, M_{\odot}$), which we focus on in this work. As a result, it is unlikely that Pop.~III physics would directly challenge the conclusions of this paper on the mean iron content. 

However, a more detailed Pop.~III model is likely necessary for detailed analysis of the low-metallicity tail of the iron MDF ([Fe/H] $\leq -4.0$; \citealt{Jeon+2017}) and all abundance ratios at such metallicities. A first step could be to account for the fact that Pop.~III enrichment affects multiple elements, rather than just oxygen, and implement an element-by-element floor calibrated on studies such as \citealt{Jaacks+2018, Brauer+2025a}). But a complete account of inhomogeneous pollution would require a dedicated implementation of the feedback and the chemical enrichment by individual Pop.~III stars. This is a natural extension of \textsc{inferno}, which we plan to undertake in future work to enable studies like the one performed here, but varying key assumptions for Pop.~III chemical enrichment in a lower-mass system. 

Interestingly, while not modeled in our simulations, we note that signatures often attributed to Pop.~III chemistry appear in our simulations. This includes metal-poor stars that are rich in carbon and nitrogen. We find several examples of this in Figures~\ref{fig:abundance_nugrid}, \ref{fig:abundance_rotation} and \ref{fig:abundance_scatter}. Furthermore, the model without SNIa displays two separate trends in [C/Fe] and [N/Fe], one at almost constant ratio and one evolving toward lower abundance with increasing [Fe/H]. We find that many of these signatures arise due to enrichment from AGB stars in low-mass proto-galaxies that are accreted later. Recent work has attributed such a signature to AGB enrichment \citep{Gil-Pons+2025}. Fully explaining their origin is beyond the scope of this work, but it is planned for a future publication (Andersson et al., in prep.).

%---------------------------------------------
\section{Conclusions}\label{sec:conclusion}

We present a suite of zoomed cosmological hydrodynamical simulations of a single UFD using an extension of the star-by-star \textsc{inferno} model. This extension generalizes the previous implementation to include detailed chemical ejecta as part of feedback from individual stars using any pre-tabulated yield table and combination of elements.

Leveraging this, we re-simulate the same dwarf galaxy 13 times, varying the input yield tables for massive stars, the yields and timescales for SNeIa, and quantifying the intrinsic stochasticity induced by IMF sampling and galaxy formation chaoticity within two chemical models. Our conclusions are: 

\begin{enumerate}
    \item SNIa are important contributors to the iron content of UFD stars. Delaying their onset from $t_{\rm Ia}=38\,{\rm Myr}$ to $t_{\rm Ia}=100\,{\rm Myr}$ systematically decreases [Fe/H] by $\sim0.4$ dex, and neglecting these explosions places galaxies well outside the luminosity-[Fe/H] relation.
    \item Stochastic sampling of the IMF and feedback timescales introduces significant galaxy-to-galaxy scatter in stellar abundances. All elements in our simulations (C, N, O, Mg, Al, Si, and Fe) are susceptible to this effect.
    \item Changing assumptions for massive star yields (e.g., including stellar rotation) result in quantitative, but not qualitative differences in mean stellar abundances. These differences between models are of the same magnitude as those induced by pure stochasticity. 
    \item However, the choice of massive star yields affects the distribution of abundance trends when viewed as a function of [Fe/H]. In particular, we find that yields from \citetalias{LimongiChieffi2018} produce bimodalities in the anti-correlation between light element abundances and [Fe/H], while others do not. 
\end{enumerate}

Our findings that mean dwarf galaxy abundances and the luminosity-[Fe/H] relation are converged with respect to different choices of yields extend the earlier findings from \citet{Buck+2021} to much lower galaxy masses. This is particularly important for studies that aim to leverage the low-mass end of the luminosity-[Fe/H] relation to constrain the implementation of feedback and its ability to drive galactic winds \citep{Wheeler+2019, Agertz+2020, Prgomet+2022, Collins&Read2022, Ibrahim+2024, Rey+2025}. 

Our results also emphasize the importance of SNIa enrichment in UFDs. In fact, the largest shifts observed in mean abundances in our results are related to variations in SNeIa parameters. Despite the importance of SNeIa modeling in UFDs \citep[see, e.g.,][]{Kirby+2019, Alexander+2025}, these are not always considered in the context of galactic chemical enrichment of faint dwarfs due to the short star formation histories of these objects and the relative rarity of SNeIa per stellar mass (\citealt{Ko+2025}). However, even in reionization-limited UFDs, star formation histories extend over $\leq 2$ Gyr, a timescale over which SNeIa easily play a role in enrichment. Furthermore, with a stellar mass of $10^5 \, M_{\odot}$, one expects $\approx 150$ SNeIa explosions, a significant fraction of which to be sampled within the first few $100\,{\rm Myr}$ (see Figure~\ref{fig:stochasticity} and discussion in Section~\ref{sec:numerical_method}). These results strongly stress the importance of considering SNeIa when modeling the chemical enrichment of even UFDs. 

Related to this, our simulations showcase a [$\alpha$/Fe] downturn related to the time when SNeIa starts contributing to Fe production. This occurs at [Fe/H]$=-2.5$ in our simulations, which is comparable to that often found in UFDs \citep{Frebel+2014, Chiti+2023}, although counter examples exist \citet{Frebel+2012}. Maybe unsurprisingly, our simulation with a galaxy of lower stellar mass (GM3: Latest, presented in \citealt{Andersson+2025}, and included in our Figure~\ref{fig:FeH_MV}) does not show a clear downturn, further highlighting a dependence on stellar mass. We explore this in detail in Andersson et al. (in prep.). Simulations from \citet{Ko+2025} report that SNIa are unable to produce the downturn in [$\alpha$/Fe] that we find at [Fe/H]$\approx-2.5$. This could be a result of their lower stellar masses reported. However, we stress that the presence of a knee depends on the massive star yields applied in our models, warranting more exploration of whether the lack of knee could be caused by their choice of CCSN yield set (\citealt{Portinari+1998}). 

Our results also emphasize the important role of stochasticity when interpreting the chemical content of a small number of dwarf galaxies. Different realizations of the same initial conditions and model can introduce differences comparable to those introduced by changing element yields. This challenge has been previously mentioned in the literature. Using simulations of dwarf galaxies from the FIRE model (\citealt{Hopkins+2014}, see also \citealt{Wheeler+2019}) with yields from \textsc{NuGrid}, \citet{Muley+2021} found a significant increase in abundance scatter (particularly [$\alpha$/Fe]) when including mass and metallicity dependent yields. The IMF sampling method also plays a role \citep{Revaz+2016}, especially for extremely low-mass systems \citep{Jeon&Ko2024}. Each of these stochastic events then compounds over time in a chaotic galaxy formation system, ultimately leading to different star formation and chemical enrichment histories \citep[see also,][]{Keller+2019, Genel+2019}.

This is an important consideration for studies aiming to constrain low-metallicity chemical enrichment models with UFD data. For single systems, random sampling of stellar masses, feedback timing, and enrichment histories can produce abundance patterns that differ by orders of magnitude, masking underlying physical correlations. These fluctuations will likely average out when considering large enough ensembles of galaxies, however, particularly since assumptions in chemical enrichment are likely to lead to systematic effects (e.g., delaying the onset of SNeIa would reduce the iron content of all dwarfs). Robust inference will require averaging over populations of dwarfs to overcome this intrinsic stochasticity. Furthermore, in addition to the galaxy-to-galaxy stochasticity, additional measurement noise is introduced by the limited number of individual stars per galaxy for which abundance measurements are made. This sampling noise can be as significant as some of the effects reported here, particularly in low-mass objects with few observable stars (\citealt{Andersson+2025}).

Combined, these two factors underscore the need for a large sample of galaxies \textit{and} large samples of stars per galaxy, a task that forthcoming spectroscopic surveys are currently undertaking \citep{Simon2019, Skuladottir+2023, Luna+2025}. Star-by-star models, which enable a direct connection to resolved star observations, combined with parameter explorations like the one performed in this study, are the perfect tools to estimate the necessary numbers of galaxies and stars per galaxy to obtain robust constraints on low-metallicity chemical enrichment assumptions. 

\begin{acknowledgements}
    The authors thank Marco Limongi, Kira Lund, Stacy Kim, Pilar Gil Pons and Brian Schmidt for insightful discussions during the buildup of this work and comments on an earlier version of the manuscript. 
    
    EPA and M-MML acknowledges support from NASA ATP grant 80NSSC24K0935 and NSF grant AST23-0795.

    JIR would like to acknowledge support from STFC grants ST/Y002865/1 and ST/Y002857/1. 

    OA acknowledges support from the Knut and Alice Wallenberg Foundation, the Swedish National Space Agency (SNSA Dnr 2023-00164), and the LMK foundation.

    A.P.J. acknowledges support from the National Science Foundation under grant AST-2307599 and the Alfred P. Sloan Foundation.

    J.M. acknowledges support from the NSF Graduate Research Fellowship Program through grant DGE-2036197.

    K.B. is supported by an NSF Astronomy and Astrophysics Postdoctoral Fellowship under award AST-2303858.
    
    This work made extensive use of the dp191 and dp324 projects on the STFC-DiRAC ecosystem. This work was performed using the DiRAC Data Intensive service at Leicester, operated by the University of Leicester IT Services, which forms part of the STFC DiRAC HPC Facility (www.dirac.ac.uk). The equipment was funded by BEIS capital funding via STFC capital grants ST/K000373/1, ST/R002363/1, and STFC DiRAC Operations grant ST/R001014/1. 

\end{acknowledgements}

\bibliographystyle{aa}
\bibliography{ref}

\begin{appendix}

\section{Metallicity floor}\label{sec:metal_floor}

\begin{figure}
    \centering
    \includegraphics[width=\linewidth]{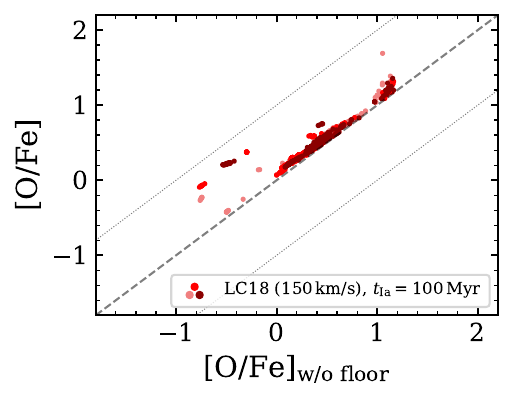}
    \caption{Comparison between the abundance of two identical tracers of oxygen, one of which (vertical axis) includes an initial floor of $2\times10^{-5}$ to promote gas cooling at early times. The figure includes data from three simulations with different seed for random number generation but the same model; yields from \citetalias{LimongiChieffi2018}, $150\,{\rm km\,s}^{-1}$ for massive stars and $t_{\rm Ia}=100\,{\rm Myr}$).}
    \label{fig:ofe_floor}
\end{figure}

To seed cooling in the gas at early times, we include a floor of $2\times10^{-5}$ in the oxygen tracer of all our simulations. Oxygen is used to compute the metallicity passed to the cooling routine. In our simulations \citetalias{LimongiChieffi2018} (150 km/s), $t_{\rm Ia}=100\,{\rm Myr}$, we included an additional oxygen tracer to test whether this initial value significantly affects the abundance that we derive. We compare the stellar abundances of these two tracers in Figure~\ref{fig:ofe_floor}, showing three examples of the same model. As expected, the [O/Fe] is always higher in the tracer with an initial value. The difference is small for [O/Fe] excluding the floor between 0 and 1, but notably, there are some outliers at low [O/Fe], offset by almost 1 dex. These points correspond to metallicities around [Fe/H]$\approx-2$. Notably, there is scatter between the simulations at these values. We also find that at the highest values, [O/Fe] diverges from the same tracer without the metal floor. In this case, there is a clear trend with [Fe/H]. The tracer without an initial value has a constant [O/Fe] value, which corresponds to that set by the ratio in the yield tables. The constant oxygen floor results in [O/Fe] shifting towards lower values as the relative amount of iron increases.

\end{appendix}

\end{document}